\documentclass[a4paper]{jpconf}
\usepackage{graphicx}
\begin{document}
\title{Underground physics with DUNE}

\author{Vitaly A. Kudryavtsev on behalf of the DUNE Collaboration}

\address{Department of Physics and Astronomy, University of Sheffield, Sheffield, S3 7RH, UK}

\ead{v.kudryavtsev@sheffield.ac.uk}

\begin{abstract}
The Deep Underground Neutrino Experiment (DUNE) is a project to design, construct and operate a next-generation long-baseline neutrino detector with a liquid argon (LAr) target capable also of searching for proton decay and supernova neutrinos. It is a merger of previous efforts of the LBNE and LBNO collaborations, as well as other interested parties to pursue a broad programme with a staged 40-kt LAr detector at the Sanford Underground Research Facility (SURF) 1300 km from Fermilab. This programme includes studies of neutrino oscillations with a powerful neutrino beam from Fermilab, as well as proton decay and supernova neutrino burst searches. In this paper we will focus on the underground physics with DUNE.
\end{abstract}

\section{The DUNE experiment}

Scientific goals of the DUNE project encompass a broad range of activities with a primary focus on neutrino oscillation studies leading to determining the neutrino mass ordering and to the measurements of CP violating phase. An underground location of the DUNE Far Detector (FD) also allows undertaking searches for new phenomena and achieving a better understanding of some astrophysical processes involving neutrino emission. Current theories beyond the Standard Model suggest that the three fundamental physical forces observed today (electromagnetic, weak and strong) were unified into one force at the beginning of the Universe. Grand Unified Theories (GUTs), which attempt to describe the unification of these forces and explain the matter-antimatter asymmetry of the Universe, predict that protons should decay, a process that has not been observed yet. DUNE will search for proton decay in the range of proton lifetimes predicted by a wide range of GUT models \cite{CDR,LBNE-so}. LAr experiments are particularly sensitive to proton decay modes, favoured by SUSY models, that involve $K^{+}$ in the final state.

DUNE will be able to detect the neutrino bursts from core-collapse supernovae within our galaxy (should any occur). Measurements of the time, flavour and energy structure of the neutrino burst will be critical for understanding the dynamics of this astrophysical phenomenon, as well as providing information on neutrino properties. The DUNE FD with its LAr target will primarily be sensitive to electron neutrinos thus complementing water Cherenkov detectors and scintillators with enhanced sensitivity to electron antineutrinos. 

The ancillary programme for underground physics with DUNE includes also measurements of neutrino oscillation parameters using atmospheric neutrinos complementing the beam neutrino measurements. In addition, a number of opportunities may arise with the development and improvement of the LAr technology, comprising measurements of solar neutrinos, detecting diffuse supernova neutrino fluxes and searches for neutrinos from extra-solar astrophysical sources, such as gamma-ray bursts, active galactic nuclei etc.

The DUNE FD will have 4 similar modules (to allow stage approach to construction) located at 4850~ft level at SURF. The muon flux is about $5.7 \times 10^{-9}$~cm$^{-2}$~s$^{-1}$ and $<E_{\mu}>\approx283$~GeV at the detector site. Each module will contain 17.1~kt of LAr, 13.8~kt of active LAr within the time projection chambers (TPCs)  and 11.6 kt fiducial mass \cite{CDR}. 

Table \ref{table:summary} summarises the main features of events relevant to underground physics with DUNE.

\begin{table}[!t]
\caption{\label{table:summary} Main features of events relevant to underground physics with the DUNE FD.}
\begin{center}
\begin{tabular}{llll}
\br
Physics & Energy range & Rate, kt$^{-1}$ year$^{-1}$ & Comments\\
\mr
Proton decay & hundreds MeV & unknown & known background \\
Atmospheric neutrinos & 0.1 -- 100 GeV & $\sim 120$ & known background \\
Supernova neutrino burst & 5 -- 50 MeV & $\sim 100$, $\Delta t \approx 10$ s, 10 kpc & known background \\
Solar neutrinos & 5 -- 15 MeV & $\sim 1300$ & high background \\
Diffuse SN neutrinos & 10 -- 50 MeV & $< 0.06$ & high background \\
\br
\end{tabular}
\end{center}
\end{table}

\section{Proton decay search}

Searches for proton decay, bound-neutron decay and neutron-antineutron oscillations test the law of conservation of baryon number.
The uniqueness of a LAr technology lies in its potential to accurately reconstruct events and particle types in the TPCs.
Electromagnetic showers are readily measured, and those from photons originated from $\pi^{0}$ decay can be distinguished to a significant degree from those coming from atmospheric $\nu_{e}$ charged-current (CC) interactions.

The proton decay mode $p \rightarrow e^{+} \pi^{0}$ is often predicted to have a high branching ratio and will give a distinct signature in all types of detector. DUNE will be able to detect this mode but is unlikely to compete on a reasonable time-scale with water Cherenkov experiments, such as Super-Kamiokande and Hyper-Kamiokande.
Another key mode is $p \rightarrow K^{+} \bar{\nu}$. This mode is dominant in most supersymmetric GUTs, many of which also favour other modes involving kaons in the final state. The decay modes with a charged kaon are unique for LAr experiments; since stopping kaons have a higher ionisation density than pions or muons with the same momentum, a LArTPC could detect them with extremely high efficiency. In addition, many final states of $K^{+}$ decay would be fully reconstructable in a LArTPC.
Table \ref{table:pdecay} summarises the efficiencies and background event rates expected in LAr for some proton decay modes. Super-Kamiokande currently has about 10\% efficiency in detecting a decay $p \rightarrow K^{+} \bar{\nu}$ \cite{SK1}.

\begin{table}[!b]
\caption{\label{table:pdecay} Efficiencies and background event rates in DUNE for some modes of proton decay.}
\begin{center}
\begin{tabular}{llllll}
\br
Mode & $p \rightarrow K^{+} \bar{\nu}$ & $p \rightarrow K^{0} \mu^{+}$ & $p \rightarrow K^{+} \mu^{-}\pi^{+}$ 
& $n \rightarrow K^{+} e^{-}$ & $n \rightarrow \e^{+} \pi^{-}$ \\
\mr
 Efficiency & 97\% & 47\% & 97\% & 96\% & 44\% \\
 Background, Mt$^{-1}$ y$^{-1}$ & 1 & $<2 $ & 1 & $<2 $ & 0.8 \\
\br
\end{tabular}
\end{center}
\end{table}

The key signature for $p \rightarrow K^{+} \bar{\nu}$ is the presence of an isolated charged kaon (which would also be monochromatic for the case of free protons, with the momentum $p \approx 340$ MeV). Unlike the case of $p \rightarrow e^{+} \pi^{0}$, where the maximum detection efficiency is limited to 40--45\% because of inelastic intra-nuclear scattering of the $\pi^{0}$, the kaon emerges intact from the nucleus with 97\% probability. The kaon momentum is smeared by the proton's Fermi motion and shifted downward by re-scattering  \cite{nuclear-effects}. The kaon emerging from this process is below Cherenkov threshold in water, therefore a water Cherenkov detector would need to detect it after it stops, via its decay products. 
In LArTPC detectors, the $K^{+}$ can be tracked and identified via detailed analysis of its energy loss profile, and its momentum measured by range. Additionally, all decay modes can be cleanly reconstructed and identified, including those with neutrinos, since the decaying proton is essentially at rest. With this level of detail, it is possible for a single event with an isolated kaon of the right momentum originating from a point within the fiducial volume, to provide overwhelming evidence for an observation of a proton decay.

\begin{figure}[htb]
\begin{minipage}{5cm}
\includegraphics[width=5cm]{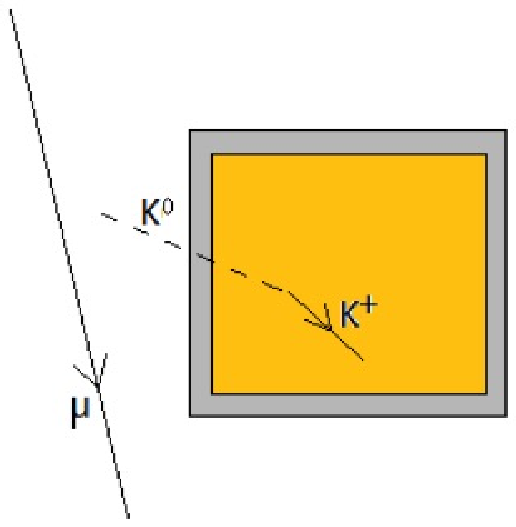}
\caption{\label{fig:background}A diagram showing a potential background event with an isolated positive kaon in a detector.}
\end{minipage}\hspace{1.5cm}%
\begin{minipage}{9cm}
\includegraphics[width=9cm]{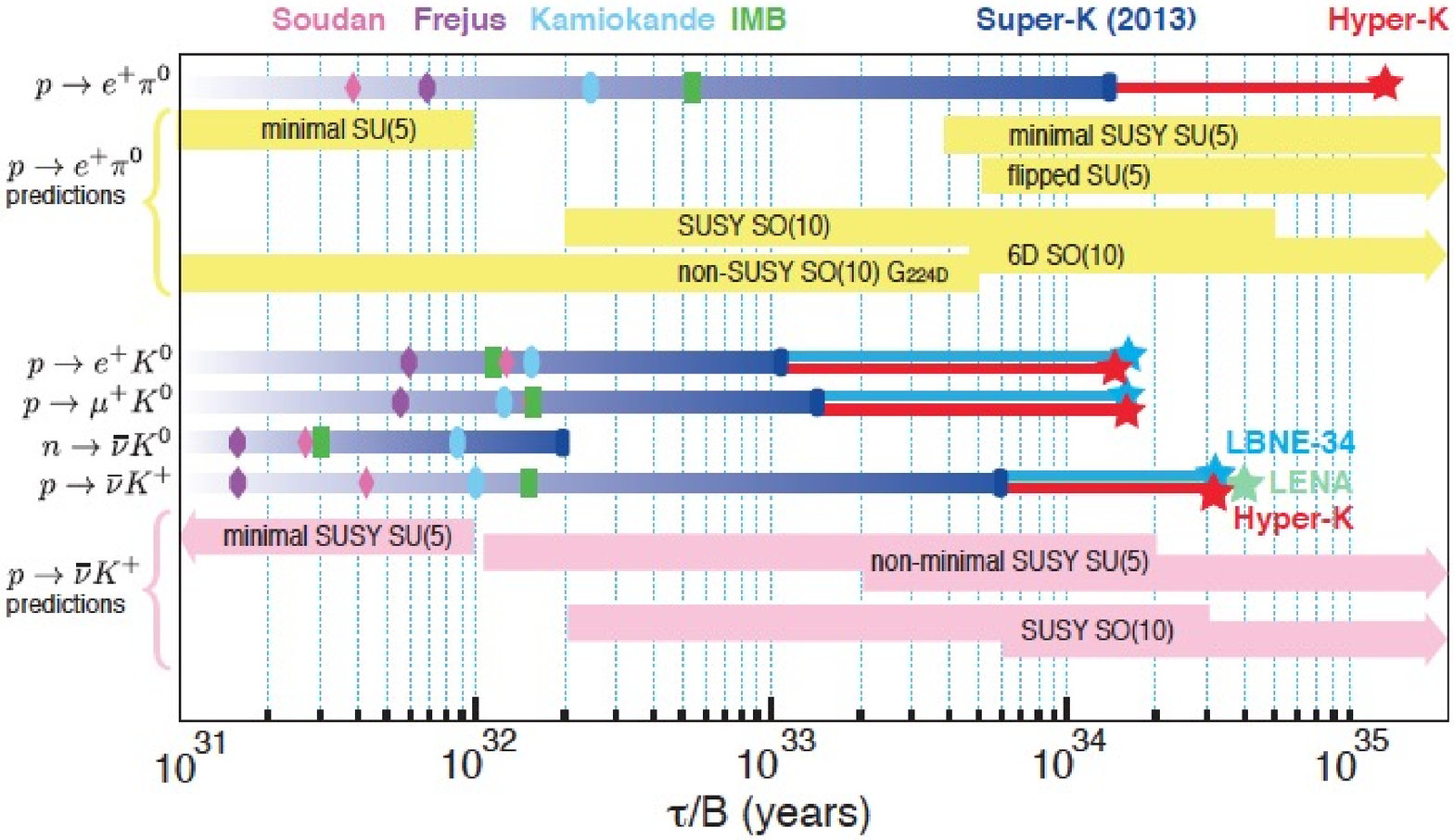}
\caption{\label{fig:sensitivity}Sensitivity of different experiments to proton lifetime for different decay modes. A 400 kt$\times$years of LAr in DUNE/LBNE is shown. For Super-Kamiokande, the current limits are shown \cite{SK,SK1}.}
\end{minipage} 
\end{figure}

The background for a proton decay search comes from atmospheric muons and neutrinos. If a muon passes through the detector, then the event can easily be reconstructed as a background muon-induced event and rejected. The current design of DUNE far detector suggests that the rate of muons passing through active LAr of a single module is about 0.05 s$^{-1}$. With a full duration of event, limited by the maximum drift time of a few ms, rejecting all events with muons crossing the active volume of the TPC will result in less than 0.1\% of dead time. Thus, only events with muons not crossing the active LAr volume may contribute to the background. Among these events the main danger comes from neutral kaons produced outside the TPC and undergoing inelastic scattering (primarily charge exchange) inside the TPC resulting in a single positive kaon detected. Figure \ref{fig:background} shows a schematic for such background event.

We have carried out initial simulations of muon-induced background in a LAr detector located at a depth of 4 km w. e. \cite{muon-background}. The following cuts have been applied to reject muon-induced events: (i) no muon is in the detector active volume, (ii) the $K^{+}$ is fully contained within the fiducial volume ($>10$~cm from any TPC wall), (iii) the energy deposition from the $K^{+}$ and its descendants (excluding decay products) is $<250$~MeV, (iv) the total energy deposition from the $K^{+}$, its descendants and decay products is $<1$~GeV, (v) energy deposition from other particles in the muon-induced cascade (i.e. excluding the energy deposition from the positive kaon, its descendants and decay products) is $<50$~MeV. No event survived the above cuts, resulting in an upper limit (at 90\% CL) on the muon-induced background rate of 0.0012 events/kt/year. 

Atmospheric neutrino background does not depend on the depth and cannot be controlled by reducing the fiducial volume. The neutrino flux is concentrated around the energy range relevant for proton decay. 
Both neutral (NC) and charge current (CC) events can contribute to this background. NC events include associated production of a pair of strange hadrons, for instance a positive kaon and a neutral baryon. In a LAr detector, the final state particles from the baryon decay should be detectable, providing a good rejection power. Events where a $K_L^{0}$ is produced together with a charged kaon but escapes detection, are more rare and can be eliminated by reducing the fiducial volume of LAr. CC events result in a change of strangeness and are suppressed by the Cabibbo angle. A charged lepton will also be produced in such event which can easily be identified.
Our initial simulations and estimates, as well as the analysis reported in Ref. \cite{Bueno}, led to the estimated background rates given in Table \ref{table:pdecay}.

Misidentification of pions in atmospheric neutrino events is a potential problem that can be controlled by measuring the residual range dependence of the particle energy deposition near the end of its trajectory (nominally 14 cm) and by the explicit reconstruction of its decay products. Also the muon momentum from a pion decay (about 30 MeV) and hence the range of the muon would not match that of the corresponding muon (236 MeV) in a kaon decay.

Figure \ref{fig:sensitivity} shows the expected sensitivity of different experiments to nucleon decay \cite{CDR}.

\section{Supernova neutrino burst}

The core-collapse neutrino signal starts with a $\sim$10~ms neutronization burst primarily composed of $\nu_e$ followed by an accretion phase lasting some hundreds of milliseconds, as matter falls onto
the collapsed core. The later cooling phase over 10 s represents the main part of the signal. Neutrino signal in LAr is detected primarily via $\nu_e + ^{40}$Ar $\rightarrow e^{-} + ^{40}$K$^{*}$ thus complementing water Cherenkov and scintillator detectors with enhanced sensitivity to electron antineutrinos. The predicted event rate from a supernova-neutrino burst may be calculated by folding expected neutrino differential energy spectra with cross sections for the relevant channels, and with detector response. For event rate estimates in liquid argon, a detection threshold of 5 MeV has been assumed. The photon-detection system of the DUNE far detector, coupled with charge collection and simple pattern recognition, is expected to provide a highly efficient trigger.

\begin{figure}[htb]
\begin{center}
\includegraphics[width=15cm]{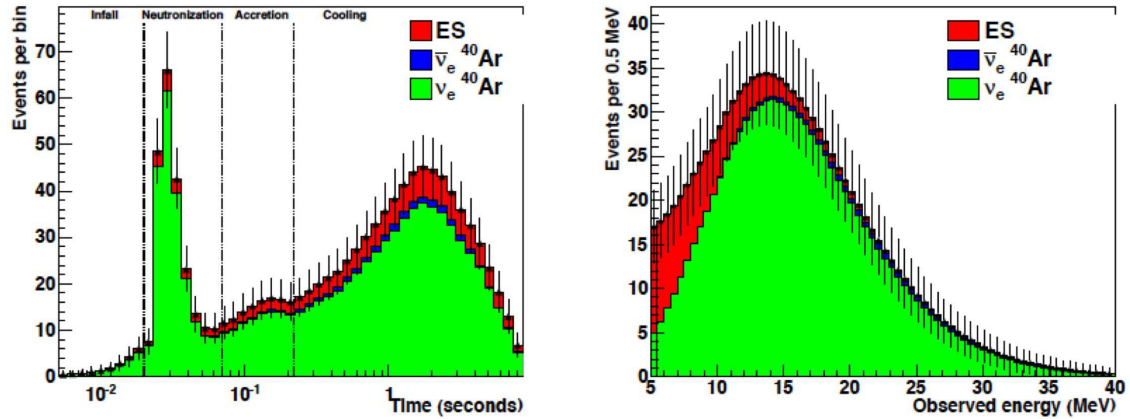}
\caption{\label{fig:snb-spectra}Left: expected time profile of the signal in 40 kt of LAr for a
supernova at 10 kpc \cite{sn-spectra}. Right: expected measured event spectrum for the same model, integrated over time.}
\end{center}
\end{figure}

Table \ref{table:snb-rates} \cite{CDR} shows rates calculated with SNOwGLoBES \cite{SNOwGLoBES} for the dominant interactions in argon for the Livermore model \cite{Livermore}, and the GKVM model \cite{GKVM}. No oscillations were assumed although oscillations will have a large, model-dependent effect. The event rate scales linearly with a fiducial mass and inverse proportionally to the square of the distance. Figure \ref{fig:snb-spectra} shows the time profile and energy spectra of events from a supernova at a distance of 10 kpc.

\begin{table}[htb]
\caption{\label{table:snb-rates} Event rates for different supernova models in 40 kt of LAr for a supernova at 10 kpc. }
\begin{center}
\begin{tabular}{lll}
\br
Reaction & Livermore model \cite{Livermore} & GKVM model \cite{GKVM} \\
\mr
$\nu_e + ^{40}$Ar $ \rightarrow e^{-} + ^{40}$K$^{*}$ & 2720 & 3350 \\
$\bar{\nu_e} + ^{40}$Ar $ \rightarrow e^{+} + ^{40}$Cl$^{*}$ & 230 & 160 \\
$\nu_x + e^{-} \rightarrow \nu_x + e^{-} $ & 350 & 260 \\
\mr
Total & 3300 & 3770 \\
\br
\end{tabular}
\end{center}
\end{table}

The background consists of  (i) intrinsic radioactive contamination in LAr or support materials, in particular $^{39}$Ar, (ii) cosmogenic radioactivity produced in situ from cosmic-rays interacting with LAr or the support materials. This background will be studied by simulations and measurements ensuring discrimination of the supernova burst signal from the background.

\section{Atmospheric neutrinos}

Atmospheric neutrinos come in all flavours and in a wide range of baseline, neutrino energy, and hence $L/E$. In principle, all oscillation parameters could be measured, with high complementarity to measurements performed with a neutrino beam. In addition, atmospheric neutrinos are available all the time, in particular before the beam becomes operational. When neutrinos travel through the Earth, the MSW resonance influences electron neutrinos in the few-GeV energy range. More precisely, the resonance occurs for $\nu_e$ in the case of normal mass hierarchy ($\Delta m^2_{32} > 0$), and for $\bar{\nu}_e$ in the case of inverted mass hierarchy ($\Delta m^2_{32} < 0$). The mass hierarchy (MH) sensitivity can be greatly enhanced if neutrino and antineutrino events can be separated. The DUNE detector will not be magnetized; however, its high-resolution imaging offers possibilities for tagging features of events that provide statistical discrimination between neutrinos and antineutrinos, for instance, a proton tag (for $\nu_e$ events) and a decaying muon tag (for $\bar{\nu}_e$). Unlike for beam measurements, the sensitivity to MH with atmospheric neutrinos is nearly independent of the CP-violating phase, thus allowing to lift degeneracies that can be present in beam analyses. Atmospheric neutrinos may also help with searching for new physics scenarios, such as CPT violation, non-standard interactions, sterile neutrinos etc.

\section{Conclusions}
A long-term operation of DUNE provides a unique opportunity to study not only neutrino oscillations with the beam neutrinos but also a large range of non-accelerator physics topics, such as atmospheric neutrinos (complementing beam neutrino measurements), proton decay search and supernova neutrinos bursts.


\section*{References}
\begin{thebibliography}{9}

\bibitem{CDR}
The DUNE Collaboration 2015 Long-Baseline Neutrino Facility (LBNF) and Deep Underground Neutrino Experiment (DUNE) Conceptual Design Report,  Vol. 2
({\it Preprint} physics.ins-det/1512.06148v1).


\bibitem{LBNE-so}
The LBNE Collaboration 2014 The Long-Baseline Neutrino Experiment. Exploring Fundamental Symmetries of the Universe ({\it Preprint} hep-ex/1307.7335v3).

\bibitem{SK1}
Abe K et al (Super-Kamiokande Collaboration) 2014 {\it Phys. Rev.} D {\bf 90} 072005.

\bibitem{nuclear-effects}
Stefan D and Ankowski A M  2009 {\it Acta Phys. Polon.} B {\bf 40} 671 ({\it Preprint} 0811.1892).

\bibitem{muon-background}
Klinger J, Kudryavtsev V A, Richardson M and Spooner N J C 2015 {\it Phys. Lett.} B {\bf 746} 44.

\bibitem{Bueno}
Bueno A et al 2007 {\it J. of High Energy Phys.} {\bf JHEP0704} 041.

\bibitem{SK}
Raaf J L for the Super-Kamiokande Collaboration 2012  {\it Nucl. Phys. Proc. Suppl.} {\bf 229-232} 559.

\bibitem{SNOwGLoBES}
Scholberg K et al, SNOwGLoBES: http://www.phy.duke.edu/$\sim$schol/snowglobes.

\bibitem{Livermore}
Totani T, Sato K, Dalhed H E and Wilson J R 1998 {\it Astrophys. J.} {\bf 496} 216.

\bibitem{GKVM}
Gava J, Kneller J, Volpe C and McLaughlin G C 2009 {\it Phys. Rev. Lett.} {\bf 103} 071101.

\bibitem{sn-spectra}
Hudepohl L, Muller B, Janka H-T, Marek A and Raffelt G 2010 {\it Phys. Rev. Lett.} {\bf 104} 251101.

\end{thebibliography}
\end{document}